\documentstyle[12pt]{article}

\begin{document}

\author{C. Bizdadea\thanks{%
e-mail address: bizdadea@hotmail.com}, I. Negru and S. O. Saliu\thanks{%
e-mail addresses: osaliu@central.ucv.ro or odile\_saliu@hotmail.com} \\
Department of Physics, University of Craiova\\
13 A. I. Cuza Str., Craiova RO-1100, Romania}
\title{Irreducible Hamiltonian BRST analysis of Stueckelberg coupled
$p$-form gauge theories}
\maketitle

\begin{abstract}
The irreducible Hamiltonian BRST symmetry for $p$-form gauge theories with
Stueckelberg coupling is derived. The cornerstone of our approach is
represented by the construction of an irreducible theory that is equivalent
from the point of view of the BRST formalism with the original system. The
equivalence makes permissible the substitution of the BRST quantization of
the reducible model by that of the irreducible theory. Our procedure
maintains the Lorentz covariance of the irreducible path integral.

\end{abstract}

\section{Introduction}

It is well-known that the Hamiltonian BRST formalism \cite{1}--\cite{5}
stands for one of the strongest and most popular quantization methods for
gauge theories. In the irreducible context the ghosts can be interpreted
like one-forms dual to the vector fields corresponding to the first-class
constraints. This geometrical interpretation fails within the reducible
framework due to the fact that the vector fields are no longer independent,
hence they cannot form a basis. The redundant behaviour generates the
appearance of ghosts with ghost number greater than one, traditionally
called ghosts for ghosts, of their canonical conjugated momenta, named
antighosts, and, in the meantime, of a pyramidal non-minimal sector. The
former objects, namely, the ghosts for ghosts, ensure a straightforward
incorporation of the reducibility relations within the cohomology of the
exterior derivative along the gauge orbits, while their antighosts are
required in order to kill the higher resolution degree non trivial co-cycles
from the homology of the Koszul-Tate differential.

A representative class of redundant systems is given by $p$-form gauge
theories, that play an important role in string and superstring theory,
supergravity and the gauge theory of gravity \cite{14}--\cite{sgrav1},
attracting much attention lately on behalf of some interesting aspects, like
their characteristic cohomology \cite{20} or their applications to higher
dimensional bosonisation \cite{21}. The study of theoretical models with $p$%
-form gauge fields give an example of so-called `topological field theory'
and lead to the appearance of topological invariants, being thus in close
relation to space-time topology, hence with lower dimensional quantum
gravity \cite{topol}. In the meantime, antisymmetric tensor fields of
various orders are included within the supergravity multiplets of many
supergravity theories \cite{sgr}, especially in 10 or 11 dimensions. It is
known that the $d=11$ supergravity is regarded as a sector of $M$-theory
unification. Of the many special properties of $d=11$ supergravity the most
interesting is that it forbids a cosmological term extension in the context
of lower-dimensional structures due precisely to the 4-form or `dual' 7-form
necessary to balance the degrees of freedom \cite{sgrav}. The construction
of `dual' Lagrangians involving $p$-forms is appears naturally in General
Relativity and supergravity in order to render manifest the ${\rm SL}(2,{\bf %
R})$ symmetry group of stationary solutions of Einstein's vacuum equation,
respectively to reveal some subtleties of `exact solutions' for supergravity 
\cite{sgrav1}. Interacting $p$-form gauge theories have been analyzed from
the redundant Hamiltonian BRST point of view in \cite{22}, where the ghost
and auxiliary field structures required by the antifield BRST formalism are
derived.

The purpose of this paper is to give a general irreducible approach to $p$%
-form gauge theories with Stueckelberg coupling in the Hamiltonian
framework. This problem is solved mainly by replacing the original redundant
Hamiltonian first-class system by an irreducible one, and by further
quantizing the resulting theory in the Hamiltonian BRST context. The
irreducible first-class system is obtained by imposing that all the
antighost number one co-cycles of the reducible Koszul-Tate differential
should identically vanish under a suitable redefinition of the antighosts at
antighost number one, and also by requesting that the number of physical
degrees of freedom of the irreducible theory to be equal with that of the
original reducible system. Initially, we analyze quadratic $p$-form gauge
theories with Stueckelberg coupling \cite{24}, and then extend the results
to the interacting case. The motivation for analyzing the Stueckelberg
coupling and not the simpler case of free abelian $p$-forms is twofold.
First, the Stueckelberg coupling is involved with the quantization of
massive $p$-forms \cite{24}, and second, this class of models presents
non-diagonal reducibility matrices as opposed to the free situation, which
makes their irreducible treatment more interesting from the quantization
point of view. Moreover, the free case can be obtained from the Stueckelberg
coupling in the limit $M=0$ (see (\ref{s1})). The idea of converting a
reducible Hamiltonian first-class theory into an irreducible one appears in 
\cite{5} and \cite{23}, but it has not been either consistently developed or
applied so far to the quantization of reducible first-class Hamiltonian
systems.

Our paper is structured in four sections. Section 2 is dealing with the
derivation of an irreducible Hamiltonian first-class system corresponding to
the starting quadratic $p$-form gauge theory with Stueckelberg coupling on
account of homological arguments, emphasizing that we can substitute the
Hamiltonian BRST quantization of the original redundant model by that of the
irreducible theory. In the end of this section we infer the path integral
for the irreducible system in the context of the Hamiltonian BRST
quantization. Section 3 investigates the extension of the analysis from
Section 2 to the interacting case. In view of this, we propose a model of
irreducible Hamiltonian theory associated to that in Section 2, and further
determine its Lagrangian version, which relies on the original action and
some Lorentz covariant irreducible gauge transformations. The interacting
case is then addressed employing the above mentioned Lagrangian version as
an appropriate starting point. Section 4 ends the paper with some
conclusions.

\section{Irreducible analysis of abelian $p$- and $\left( p-1\right) $-forms
with Stueckelberg coupling}

In this section we derive the path integral of abelian $p$- and $\left(
p-1\right) $-forms with Stueckelberg coupling following an irreducible
approach. Thus, we begin with the canonical analysis of the model, which is
described by a quadratic Lagrangian action and displays a $\left( p-1\right) 
$-stage reducible first-class constraint set. In subsection 2.2 we construct
some irreducible first-class constraints associated with the original ones
by means of homological arguments. More precisely, we require that all the
non trivial antighost number one co-cycles of the reducible Koszul-Tate
differential identically vanish under a suitable redefinition of the
antighost number one antighosts while preserving the original number of
physical degrees of freedom. The analysis is performed for $p=2$ and $p=3$,
in order to emphasize various aspects of graduate complexity, which will be
employed in order to generalize our results to arbitrary $p$. In this
manner, we arrive at an irreducible set of first-class constraints, a
corresponding first-class Hamiltonian and an irreducible Koszul-Tate
differential associated with the starting reducible model. With these
elements at hand, we construct in subsection 2.3 the irreducible BRST
symmetry and show that it exists as it satisfies the general requirements of
the homological perturbation theory. In the next subsection we elucidate the
relationship between the reducible and irreducible BRST symmetries by
proving that the physical observables deriving from the two contexts
coincide, such that it is permissible to replace the Hamiltonian BRST
quantization of the original reducible model by that of the resulting
irreducible theory. In subsection 2.5 we apply the Hamiltonian BRST
formalism to the irreducible model by using a proper non-minimal sector and
a gauge-fixing fermion that finally lead to a manifestly covariant path
integral.

\subsection{Description of the model}

Our starting point is the quadratic Lagrangian action%
\begin{eqnarray}\label{s1}
& &S_0^L\left[ A_{\mu _1\ldots \mu _p},H_{\mu _1\ldots \mu _{p-1}}\right]
=-\int d^dx\left( \frac 1{2\cdot \left( p+1\right) !}F_{\mu _1\ldots \mu
_{p+1}}F^{\mu _1\ldots \mu _{p+1}}+\right. \nonumber \\ 
& &\left. \frac 1{2\cdot p!}\left( MA_{\mu _1\ldots \mu _p}-F_{\mu
_1\ldots \mu _p}\right) \left( MA^{\mu _1\ldots \mu _p}-F^{\mu _1\ldots \mu
_p}\right) \right) , 
\end{eqnarray}
where $F_{\mu _1\ldots \mu _{p+1}}$ and $F_{\mu _1\ldots \mu _p}$ represent
the field strengths of $A_{\mu _1\ldots \mu _p}$, respectively, $H_{\mu
_1\ldots \mu _{p-1}}$. Of course, it is understood that $d\geq p+1$. Action (%
\ref{s1}) is invariant under the gauge transformations 
\begin{equation}
\label{s1a}\delta _\epsilon A^{\mu _1\ldots \mu _p}=\partial ^{\left[ \mu
_1\right. }\epsilon ^{\left. \mu _2\ldots \mu _p\right] },
\end{equation}
\begin{equation}
\label{s1b}\delta _\epsilon H^{\mu _1\ldots \mu _{p-1}}=\partial ^{\left[
\mu _1\right. }\bar \epsilon ^{\left. \mu _2\ldots \mu _{p-1}\right]
}+M\epsilon ^{\mu _1\ldots \mu _{p-1}},
\end{equation}
where $\left[ \mu _1\cdots \mu _k\right] $ signifies antisymmetry with
respect to the indices between brackets.

Performing the canonical analysis of (\ref{s1}), one infers the first-class
constraints 
\begin{equation}
\label{s2}G_{i_1\ldots i_{p-1}}^{(1)}\equiv \pi _{0i_1\ldots i_{p-1}}\approx
0, 
\end{equation}
\begin{equation}
\label{s3}G_{i_1\ldots i_{p-2}}^{(1)}\equiv \Pi _{0i_1\ldots i_{p-2}}\approx
0, 
\end{equation}
\begin{equation}
\label{s4}G_{i_1\ldots i_{p-1}}^{(2)}\equiv -p\partial ^i\pi _{ii_1\ldots
i_{p-1}}+M\Pi _{i_1\ldots i_{p-1}}\approx 0, 
\end{equation}
\begin{equation}
\label{s5}G_{i_1\ldots i_{p-2}}^{(2)}\equiv -\left( p-1\right) \partial
^i\Pi _{ii_1\ldots i_{p-2}}\approx 0, 
\end{equation}
and the canonical Hamiltonian%
\begin{eqnarray}\label{s6}
& &\bar H=\int d^{d-1}x\left( -\frac{p!}2\pi _{i_1\ldots i_p}\pi ^{i_1\ldots
i_p}-\frac{\left( p-1\right) !}2\Pi _{i_1\ldots i_{p-1}}\Pi ^{i_1\ldots
i_{p-1}}+\right. \nonumber \\ 
& &A^{0i_1\ldots i_{p-1}}G_{i_1\ldots i_{p-1}}^{(2)}+\frac 1{2\cdot p!}\left(
MA_{i_1\ldots i_p}-F_{i_1\ldots i_p}\right) \left( MA^{i_1\ldots
i_p}-F^{i_1\ldots i_p}\right) +\nonumber \\ 
& &\left. \frac 1{2\cdot \left( p+1\right) !}F_{i_1\ldots
i_{p+1}}F^{i_1\ldots i_{p+1}}+H^{0i_1\ldots i_{p-2}}G_{i_1\ldots
i_{p-2}}^{(2)}\right) . 
\end{eqnarray}
In (\ref{s2}--\ref{s6}) $\pi $ and $\Pi $ stand for the momenta of the
corresponding $A$, respectively, $H$. Using the notations 
\begin{equation}
\label{s7}G_{a_0}^{(2)}\equiv \left( G_{i_1\ldots
i_{p-1}}^{(2)},G_{i_1\ldots i_{p-2}}^{(2)}\right) , 
\end{equation}
we find that the constraint functions (\ref{s7}) are $\left( p-1\right) $%
-stage reducible 
\begin{equation}
\label{s8}Z_{\;\;a_1}^{a_0}G_{a_0}^{(2)}=0, 
\end{equation}
\begin{equation}
\label{s9}Z_{\;\;a_{k-1}}^{a_{k-2}}Z_{\;\;a_k}^{a_{k-1}}=0,\;k=2,\ldots
,p-1, 
\end{equation}
where the $k$th order reducibility functions are expressed by 
\begin{equation}
\label{s10}Z_{\;\;a_k}^{a_{k-1}}=\left( 
\begin{array}{cc}
\frac 1{\left( p-k-1\right) !}\partial _{}^{\left[ i_1\right. }\delta
_{\;\;j_1}^{i_2}\ldots \delta _{\;\;j_{p-k-1}}^{\left. i_{p-k}\right] } & 
{\bf 0} \\ \frac{\left( -\right) ^{k+1}M}{\left( p-k-1\right) !}\delta
_{\;\;j_1}^{\left[ i_1\right. }\ldots \delta _{\;\;j_{p-k-1}}^{\left.
i_{p-k-1}\right] } & \frac 1{\left( p-k-2\right) !}\partial _{}^{\left[
i_1\right. }\delta _{\;\;j_1}^{i_2}\ldots \delta _{\;\;j_{p-k-2}}^{\left.
i_{p-k-1}\right] } 
\end{array}
\right) , 
\end{equation}
$k=1,\ldots ,p-1$, and 
\begin{equation}
\label{s11}a_k=\left( j_1\ldots j_{p-k-1},j_1\ldots j_{p-k-2}\right)
,\;k=0,\ldots ,p-1. 
\end{equation}
Throughout the paper we work with the conventions $f^{i_1\cdots i_m}=f$ if $%
m=0$ and $f^{i_1\cdots i_m}=0$ if $m<0$. This ends the canonical analysis of
this model.

\subsection{Irreducible constraints}

The first step of our analysis consists in the derivation of an irreducible
first-class theory associated with the original reducible one, which, in
addition, preserves the initial number of physical degrees of freedom. In
view of this we construct some irreducible first-class constraints starting
with the reducible constraints (\ref{s4}--\ref{s5}). Clearly, the case $p=1$
is irreducible and in consequence it will be not discussed in the sequel.
For a deeper understanding of our procedure we initially expose the case $%
p=2 $, subsequently explore the situation $p=3$, and finally generalize our
results to an arbitrary $p$.

\subsubsection{The case $p=2$}

In this case the constraints (\ref{s4}--\ref{s5}) take the form 
\begin{equation}
\label{a1}G_i^{(2)}\equiv -2\partial ^j\pi _{ji}+M\Pi _i\approx
0,\;G^{(2)}\equiv -\partial ^i\Pi _i\approx 0, 
\end{equation}
and are first-stage reducible 
\begin{equation}
\label{a2}\partial ^iG_i^{(2)}+MG^{(2)}=0. 
\end{equation}
The reducible BRST symmetry 
\begin{equation}
\label{a3}s_R=\delta _R+\sigma _R+\cdots , 
\end{equation}
contains two basic differentials. The first one, $\delta _R$, named the
Koszul-Tate differential, realizes an homological resolution of smooth
functions defined on the constraint surface, while the second one, $\sigma
_R $, represents a model of longitudinal derivative along the gauge orbits
and accounts for the gauge invariances (generated by the first-class
constraints). In order to realize a proper construction of $\delta _R$ we
set the action of this operator on all phase-space variables to vanish and
introduce some new generators, called antighosts and denoted by ${\cal P}%
_{2i}$, $P_2$, and $\lambda $, accordingly to which we define 
\begin{equation}
\label{a4}\delta _R{\cal P}_{2i}=-G_i^{(2)},\;\delta _RP_2=-G^{(2)}, 
\end{equation}
\begin{equation}
\label{a5}\delta _R\lambda =-\partial ^i{\cal P}_{2i}-MP_2. 
\end{equation}
While ${\cal P}_{2i}$ and $P_2$ are fermionic fields of antighost number
one, the antighost $\lambda $ is bosonic and possesses antighost number two.
The antighost $\lambda $ is required in order to make the co-cycle 
\begin{equation}
\label{a6}\mu =\partial ^i{\cal P}_{2i}+MP_2, 
\end{equation}
$\delta _R$-exact, which will establish the acyclicity of the Koszul-Tate
differential. Our main idea of passing to an irreducible treatment is to
redefine the antighosts ${\cal P}_{2i}$ and $P_2$ in such a way that the new
co-cycle of the type (\ref{a6}) vanishes identically. If we implement this
step, the new co-cycle at antighost number one will be trivially
(identically vanishing) without introducing $\lambda $, hence the resulting
theory will be indeed irreducible. The redefinition of the antighosts is
performed through 
\begin{equation}
\label{a7}{\cal P}_{2i}\rightarrow \stackrel{\symbol{126}}{\cal P}%
_{2i}=D_{\;\;i}^j{\cal P}_{2j}+\tilde D_iP_2,\;P_2\rightarrow \tilde P_2=D^j%
{\cal P}_{2j}+DP_2, 
\end{equation}
where the quantities $D_{\;\;i}^j$, $\tilde D_i$, $D^j$ and $D$ are taken to
satisfy the equations 
\begin{equation}
\label{a8}\partial ^iD_{\;\;i}^j+D^j=0,\;\partial ^i\tilde D_i+MD=0, 
\end{equation}
\begin{equation}
\label{a9}D_{\;\;i}^jG_j^{(2)}+\tilde
D_iG^{(2)}=G_i^{(2)},\;D^jG_j^{(2)}+DG^{(2)}=G^{(2)}. 
\end{equation}
Taking into account (\ref{a4}), (\ref{a7}) and (\ref{a9}), after simple
computation we find that 
\begin{equation}
\label{a10}\delta \stackrel{\symbol{126}}{\cal P}_{2i}=-G_i^{(2)},\;\delta 
\stackrel{\symbol{126}}{P}_2=-G^{(2)}. 
\end{equation}
From (\ref{a10}) we obtain that the co-cycle of the type (\ref{a6}), $\tilde
\mu =\partial ^i\stackrel{\symbol{126}}{\cal P}_{2i}+M\stackrel{\symbol{126}%
}{P}_2$, vanishes identically due to (\ref{a8}). We redenoted the
Koszul-Tate differential by $\delta $ in order to emphasize that it
corresponds to an irreducible situation. In this way our scope, namely, to
make $\tilde \mu $ vanish, can be attained if the system (\ref{a8}--\ref{a9}%
) is solvable. The solution to this system exists and is given by 
\begin{equation}
\label{a12}D_{\;\;i}^j=\delta _{\;\;i}^j-\frac{\partial ^j\partial _i}{%
\triangle +M^2},\;D^j=-M\frac{\partial ^j}{\triangle +M^2}, 
\end{equation}
\begin{equation}
\label{a13}\tilde D_i=-M\frac{\partial _i}{\triangle +M^2},\;D=1-\frac{M^2}{%
\triangle +M^2}, 
\end{equation}
where $\triangle =\partial _k\partial ^k$. Replacing (\ref{a12}--\ref{a13})
in (\ref{a10}) we arrive at 
\begin{equation}
\label{a14}\delta {\cal P}_{2i}-\frac{\partial _i}{\triangle +M^2}\delta
\left( \partial ^j{\cal P}_{2j}+MP_2\right) =-G_i^{(2)}, 
\end{equation}
\begin{equation}
\label{a15}\delta P_2-\frac M{\triangle +M^2}\delta \left( \partial ^j{\cal P%
}_{2j}+MP_2\right) =-G^{(2)}. 
\end{equation}
The last relations describe the action of $\delta $ corresponding to an
irreducible model subject to some irreducible first-class constraints to be
further determined. At this point we explore the requirement that the number
of physical degrees of freedom should be preserved by passing to the
irreducible theory. As the number of independent constraint functions (\ref
{a1}) is equal to $\left( d-1\right) $ and that of independent constraint
functions implicitly involved with (\ref{a14}--\ref{a15}) is $d$, it results
that we need an extra degree of freedom for the irreducible theory. We
denote this supplementary degree of freedom by $\left( A,\pi \right) $, with 
$\pi $ the non-vanishing solution to the equation 
\begin{equation}
\label{a16}\left( \triangle +M^2\right) \pi =\delta \left( \partial ^j{\cal P%
}_{2j}+MP_2\right) . 
\end{equation}
Due to the invertibility of $\left( \triangle +M^2\right) $, the
non-vanishing solution for $\pi $ enforces the irreducibility because the
equation (\ref{a16}) possesses non-vanishing solutions if and only if $%
\delta \left( \partial ^j{\cal P}_{2j}+MP_2\right) \neq 0$, hence if and
only if (\ref{a6}) is not a co-cycle. Making use of (\ref{a14}--\ref{a16})
we get that 
\begin{equation}
\label{a17}\delta {\cal P}_{2i}=-G_i^{(2)}+\partial _i\pi ,\;\delta
P_2=-G^{(2)}+M\pi . 
\end{equation}
The above relations are nothing but the definitions of $\delta $ on the
antighost number one antighosts corresponding to an irreducible theory
subject to the irreducible first-class constraints 
\begin{equation}
\label{a18}\gamma _i^{(2)}\equiv -2\partial ^j\pi _{ji}+M\Pi _i-\partial
_i\pi \approx 0,\;T^{(2)}\equiv -\partial ^i\Pi _i-M\pi \approx 0. 
\end{equation}
This solves the problem of constructing some irreducible first-class
constraints deriving from (\ref{a1}) in the case $p=2$.

It seems that our irreducible approach gives raise to some problems linked
with locality. Indeed, $\stackrel{\symbol{126}}{\cal P}_{2i}$ and $\stackrel{%
\symbol{126}}{P}_2$, explicitly written in (\ref{a14}--\ref{a15}), contain a
non-local term. This is not a surprise as $\stackrel{\symbol{126}}{\cal P}%
_{a_0}=\left( \stackrel{\symbol{126}}{\cal P}_{2i},\stackrel{\symbol{126}}{P}%
_2\right) $ are nothing but the `transverse' part of ${\cal P}_{a_0}=\left( 
{\cal P}_{2i},P_2\right) $ with respect to $Z^{a_0}=\left( 
\begin{array}{c}
\partial ^i \\ 
M 
\end{array}
\right) $, i.e., $Z^{a_0}\stackrel{\symbol{126}}{\cal P}_{a_0}=0$, and it is
known that the decomposition into transverse and longitudinal components
generates non-locality. On the other hand, the solution of the equation (\ref
{a16}) is generally non-local. However, the non-locality of the solution to (%
\ref{a16}) compensates in a certain sense the non-locality present in (\ref
{a14}--\ref{a15}) such that the resulting irreducible constraints (\ref{a18}%
) (inferred via (\ref{a17})) are local. Anticipating a bit, in the case of $%
p\geq 3$ the redefinition of the antighosts will consequently imply some
`transverse'-type conditions with respect to the corresponding reducibility
functions which lead to some non-local solutions. In order to compensate
this non-locality and to further obtain some local irreducible constraints
it will be also necessary to add some supplementary degrees of freedom that
check some equations of the type (\ref{a16}). In general, the lack of
locality in the BRST formalism can occur if the BRST charge or the
gauge-fixing fermion are non-local. As it will be seen below, this does not
happen in the context of our procedure (see (\ref{s40}) and (\ref{s42})).

\subsubsection{The case $p=3$}

The guide line in this situation is the case $p=2$. However, we will see
that some new features arise. The starting reducible constraints have the
form 
\begin{equation}
\label{a19}G_{ij}^{(2)}\equiv -3\partial ^k\pi _{kij}+M\Pi _{ij}\approx
0,\;G_i^{(2)}\equiv -2\partial ^j\Pi _{ji}\approx 0. 
\end{equation}
The definition of the reducible Koszul-Tate differential reads as 
\begin{equation}
\label{a20}\delta _R{\cal P}_{2ij}=-G_{ij}^{(2)},\;\delta
_RP_{2i}=-G_i^{(2)}, 
\end{equation}
\begin{equation}
\label{a21}\delta _R\lambda _i=-2\partial ^j{\cal P}_{2ji}-MP_{2i},\;\delta
_R\lambda =-\partial ^iP_{2i}, 
\end{equation}
\begin{equation}
\label{a22}\delta _R\tilde \lambda =-\partial ^i\lambda _i+M\lambda , 
\end{equation}
where ${\cal P}_{2ij}$ and $P_{2i}$ are fermionic of antighost number one, $%
\lambda _i$ and $\lambda $ are bosonic with antighost number two, and $%
\tilde \lambda $ is fermionic of antighost number three. The antighosts $%
\lambda _i$ and $\lambda $ must be introduced in order to enforce the $%
\delta _R$-exactness of the antighost number one co-cycles 
\begin{equation}
\label{a23}\nu _i=2\partial ^j{\cal P}_{2ji}+MP_{2i},\;\nu =\partial
^iP_{2i}, 
\end{equation}
while the presence of $\tilde \lambda $ solves the exactness of the
antighost number two co-cycle 
\begin{equation}
\label{a24}\alpha =\partial ^i\lambda _i-M\lambda . 
\end{equation}
We apply the same idea like before, namely, we demand that $\nu _i$ and $\nu 
$ are no longer co-cycles in the irreducible context (described in terms of
the irreducible Koszul-Tate operator $\delta $), so they must be no longer $%
\delta $-closed. This request can be satisfied if we add some new bosonic
canonical pairs $\left( A^i,\pi _i\right) $, $\left( H,\Pi \right) $ and
impose that the momenta $\pi _i$ and $\Pi $ are the non-vanishing solutions
to the equations 
\begin{equation}
\label{a25}\delta \left( 2\partial ^j{\cal P}_{2ji}+MP_{2i}\right) =\left(
\triangle +M^2\right) \pi _i, 
\end{equation}
\begin{equation}
\label{a26}\delta \left( \partial ^iP_{2i}\right) =\left( \triangle
+M^2\right) \Pi . 
\end{equation}
From (\ref{a25}) we get that $\delta \left( M\partial ^iP_{2i}\right)
=\left( \triangle +M^2\right) \partial ^i\pi _i$, which combined with (\ref
{a26}) leads to 
\begin{equation}
\label{a27}\partial ^i\pi _i-M\Pi =0, 
\end{equation}
on behalf of the invertibility of $\left( \triangle +M^2\right) $. The last
relation is a new constraint of the irreducible theory that ensures the
preservation of the physical degrees of freedom with respect to the initial
model. Indeed, the number of independent constraints (\ref{a19}) is $\left(
d-1\right) \left( d-2\right) /2$. By contrast, in the irreducible framework
we will find precisely $\left( d-1\right) \left( d-2\right) /2+\left(
d-1\right) $ irreducible constraints corresponding to (\ref{a19}) and a
supplementary number of phase-space variables $\left( A^i,\pi _i\right) $, $%
\left( H,\Pi \right) $, which is equal to $2d$. Thus, in order to re-obtain
the original number of degrees of freedom, it is necessary to add an extra
constraint, which forms together with the others an irreducible first-class
set. This constraint is nevertheless offered precisely by (\ref{a27}) and
will be denoted by 
\begin{equation}
\label{a28}\gamma ^{(2)}\equiv -\partial ^i\pi _i+M\Pi \approx 0. 
\end{equation}
We remark that from the entire set of first-class constraints (\ref{a19})
and (\ref{a28}), the latter is already irreducible, so its presence does not
imply further co-cycles at antighost number one. The antighost corresponding
to (\ref{a28}), ${\cal P}_2$, is fermionic, has the antighost number equal
to one, and must satisfy 
\begin{equation}
\label{a29}\delta {\cal P}_2=-\gamma ^{(2)}. 
\end{equation}
At this stage we redefine the antighost number one antighosts in order to
make the co-cycles of the type (\ref{a23}) to vanish identically. The
redefinition reads as 
\begin{equation}
\label{a30}{\cal P}_{2ij}\rightarrow \stackrel{\symbol{126}}{\cal P}%
_{2ij}=D_{\;\;ij}^{kl}{\cal P}_{2kl}+\tilde D_{\;\;ij}^kP_{2k}, 
\end{equation}
\begin{equation}
\label{a31}P_{2i}\rightarrow \stackrel{\symbol{126}}{P}_{2i}=D_{\;\;i}^{kl}%
{\cal P}_{2kl}+D_{\;\;i}^kP_{2k}, 
\end{equation}
with $D_{\;\;ij}^{kl}$, $\tilde D_{\;\;ij}^k$, $D_{\;\;i}^{kl}$ and $%
D_{\;\;i}^k$ taken to fulfill 
\begin{equation}
\label{a32}2\partial ^iD_{\;\;ij}^{kl}+MD_{\;\;j}^{kl}=0,\;2\partial
^i\tilde D_{\;\;ij}^k+MD_{\;\;j}^k=0, 
\end{equation}
\begin{equation}
\label{a33}\partial ^iD_{\;\;i}^{kl}=0,\;\partial ^iD_{\;\;i}^k=0, 
\end{equation}
\begin{equation}
\label{a34}D_{\;\;ij}^{kl}G_{kl}^{(2)}+\tilde
D_{\;\;ij}^kG_k^{(2)}=G_{ij}^{(2)}, 
\end{equation}
\begin{equation}
\label{a35}D_{\;\;i}^{kl}G_{kl}^{(2)}+D_{\;\;i}^kG_k^{(2)}=G_i^{(2)}. 
\end{equation}
In consequence, the relations (\ref{a20}) become 
\begin{equation}
\label{a36}\delta \stackrel{\symbol{126}}{\cal P}_{2ij}=-G_{ij}^{(2)},\;%
\delta \stackrel{\symbol{126}}{P}_{2i}=-G_i^{(2)}, 
\end{equation}
which further lead to the co-cycles $\tilde \nu _i=2\partial ^j\stackrel{%
\symbol{126}}{\cal P}_{2ji}+M\stackrel{\symbol{126}}{P}_{2i}$, $\tilde \nu
=\partial ^i\stackrel{\symbol{126}}{P}_{2i}$ that vanish identically due to (%
\ref{a32}--\ref{a33}). The solution of (\ref{a32}--\ref{a35}) takes the form 
\begin{equation}
\label{a37}D_{\;\;ij}^{kl}=\frac 12\delta _{\;\;i}^{\left[ k\right. }\delta
_{\;\;j}^{\left. l\right] }-\frac 1{2\left( \triangle +M^2\right) }\delta
_{\;\;m}^{\left[ l\right. }\partial _{}^{\left. k\right] }\delta _{\left[
j\right. }^m\partial _{\left. i\right] }^{},\;\tilde D_{\;\;ij}^k=-\frac
M{2\left( \triangle +M^2\right) }\delta _{\left[ j\right. }^k\partial
_{\left. i\right] }^{}, 
\end{equation}
\begin{equation}
\label{a38}D_{\;\;i}^{kl}=-\frac M{\triangle +M^2}\delta _{\;\;i}^{\left[
l\right. }\partial _{}^{\left. k\right] },\;D_{\;\;i}^k=\left( 1-\frac{M^2}{%
\triangle +M^2}\right) \delta _{\;\;i}^k-\frac{\partial ^k\partial _i}{%
\triangle +M^2}. 
\end{equation}
Substituting (\ref{a37}--\ref{a38}) in (\ref{a36}) and taking into account (%
\ref{a25}--\ref{a26}), we finally deduce 
\begin{equation}
\label{a39}\delta {\cal P}_{2ij}=-G_{ij}^{(2)}+\frac 12\partial _{\left[
i\right. }\pi _{\left. j\right] },\;\delta P_{2i}=-G_i^{(2)}+M\pi
_i+\partial _i\Pi . 
\end{equation}
The last formulas emphasize the irreducible first-class constraints 
\begin{equation}
\label{a40}\gamma _{ij}^{(2)}\equiv -3\partial ^k\pi _{kij}+M\Pi _{ij}-\frac
12\partial _{\left[ i\right. }\pi _{\left. j\right] }\approx 0, 
\end{equation}
\begin{equation}
\label{a41}T_i^{(2)}\equiv -2\partial ^j\Pi _{ji}-M\pi _i-\partial _i\Pi
\approx 0, 
\end{equation}
which together with (\ref{a28}) form the searched for irreducible set
associated with (\ref{a19}). The new feature arising in the $p=3$ case is
given by the appearance of the new constraint (\ref{a28}). Like in the case $%
p=2$, the non-locality present in (\ref{a25}--\ref{a26}) compensates the
non-locality produced by (\ref{a37}--\ref{a38}) in (\ref{a36}). We will see
that in the more complex situations $p\geq 4$ we have to add more
constraints instead of (\ref{a28}).

\subsubsection{Generalization to arbitrary $p$}

Acting along the line exposed above, we introduce the antisymmetric
canonical pairs 
\begin{equation}
\label{s16}\left( A^{j_1\ldots j_{p-2k-2}},\pi _{j_1\ldots
j_{p-2k-2}}\right) ,\left( H^{j_1\ldots j_{p-2k-3}},\Pi _{j_1\ldots
j_{p-2k-3}}\right) , 
\end{equation}
for $k\geq 0$ in order to prevent the appearance of any antighost number one
co-cycle, and, by using some homological arguments similar with the previous
ones, we construct the tower of irreducible first-class constraints
associated with (\ref{s4}--\ref{s5}) under the form 
\begin{eqnarray}\label{s22}
& &\gamma _{i_1\ldots i_{p-2k-1}}^{(2)}\equiv -\left( p-2k\right) \partial
^i\pi _{ii_1\ldots i_{p-2k-1}}+M\Pi _{i_1\ldots i_{p-2k-1}}-\nonumber \\ 
& &\frac 1{p-2k-1}\partial _{\left[ i_1\right. }\pi _{\left.
i_2\ldots i_{p-2k-1}\right] }\approx 0,\;k=0,\ldots ,a, 
\end{eqnarray}
\begin{eqnarray}\label{s23}
& &T_{i_1\ldots i_{p-2k-2}}^{(2)}\equiv -\left( p-2k-1\right) \partial ^i\Pi
_{ii_1\ldots i_{p-2k-2}}-M\pi _{i_1\ldots i_{p-2k-2}}-\nonumber \\ 
& &\frac 1{p-2k-2}\partial _{\left[ i_1\right. }\Pi _{\left.
i_2\ldots i_{p-2k-2}\right] }\approx 0,\;k=0,\ldots ,c, 
\end{eqnarray}
where we used the notations 
\begin{equation}
\label{s23a}a=\left\{ 
\begin{array}{c}
\frac p2-1,\; 
{\rm for}\;p\;{\rm even}, \\ \frac{p-1}2,\;{\rm for}\;p\;{\rm odd}, 
\end{array}
\right. \;\;c=\left\{ 
\begin{array}{c}
\frac p2-1,\; 
{\rm for}\;p\;{\rm even}, \\ \frac{p-3}2,\;{\rm for}\;p\;{\rm odd}. 
\end{array}
\right. 
\end{equation}
At this moment the constraints of the irreducible theory are expressed by (%
\ref{s2}--\ref{s3}) and (\ref{s22}--\ref{s23}).

Due to the fact that we intend to develop a covariant irreducible approach,
it is necessary to further enlarge the phase-space by adding the
antisymmetric canonical pairs 
\begin{equation}
\label{s18}\left( A^{0j_1\ldots j_{p-2k-1}},\pi _{0j_1\ldots
j_{p-2k-1}}\right) ,\left( H^{0j_1\ldots j_{p-2k-2}},\Pi _{0j_1\ldots
j_{p-2k-2}}\right) , 
\end{equation}
for $k\geq 1$, which we impose to be constrained by 
\begin{equation}
\label{s19}\pi _{0j_1\ldots j_{p-2k-1}}\approx 0,\;\Pi _{0j_1\ldots
j_{p-2k-2}}\approx 0,\;k\geq 1. 
\end{equation}
In this manner, the constraints of the irreducible theory are expressed by (%
\ref{s22}--\ref{s23}) and also by 
\begin{equation}
\label{s20}\gamma _{i_1\ldots i_{p-2k-1}}^{(1)}\equiv \pi _{0i_1\ldots
i_{p-2k-1}}\approx 0,\;k=0,\ldots ,a, 
\end{equation}
\begin{equation}
\label{s21}T_{i_1\ldots i_{p-2k-2}}^{(1)}\equiv \Pi _{0i_1\ldots
i_{p-2k-2}}\approx 0,\;k=0,\ldots ,c, 
\end{equation}
which form a first-class, abelian and irreducible set. The first-class
Hamiltonian with respect to the above first-class constraints will be taken
of the form%
\begin{eqnarray}\label{s24}
& &\bar H=\int d^{d-1}x\left( -\frac{p!}2\pi _{i_1\ldots i_p}\pi ^{i_1\ldots
i_p}-\frac{\left( p-1\right) !}2\Pi _{i_1\ldots i_{p-1}}\Pi ^{i_1\ldots
i_{p-1}}+\right. \nonumber \\ 
& &\sum\limits_{k=0}^aA^{0i_1\ldots i_{p-2k-1}}\gamma _{i_1\ldots
i_{p-2k-1}}^{(2)}+\frac 1{2\cdot p!}\left( MA_{i_1\ldots i_p}-F_{i_1\ldots
i_p}\right) \left( MA^{i_1\ldots i_p}-F^{i_1\ldots i_p}\right) +\nonumber \\ 
& &\left. \frac 1{2\cdot \left( p+1\right) !}F_{i_1\ldots
i_{p+1}}F^{i_1\ldots i_{p+1}}+\sum\limits_{k=0}^cH^{0i_1\ldots
i_{p-2k-2}}T_{i_1\ldots i_{p-2k-2}}^{(2)}\right) . 
\end{eqnarray}
At the level of the extended action of the initial model the gauge
variations of the Lagrange multipliers for the reducible constraints contain
some supplementary gauge parameters due to the reducibility, which ensure
the covariance of the Lagrangian gauge transformations. On the contrary, in
the framework of our irreducible treatment these additional gauge parameters
are absent because of the irreducibility. This is why we cannot yet build a
model of irreducible extended formalism that outputs some covariant
Lagrangian gauge transformations. In order to restore the covariance, it is
necessary to add a number of supplementary pairs equal with double of the
number of pairs (\ref{s16}) 
\begin{equation}
\label{s25}\left( B^{(1)j_1\ldots j_{p-2k-2}},\pi _{j_1\ldots
j_{p-2k-2}}^{(1)}\right) ,\left( V^{(1)j_1\ldots j_{p-2k-3}},\Pi _{j_1\ldots
j_{p-2k-3}}^{(1)}\right) , 
\end{equation}
\begin{equation}
\label{s26}\left( B^{(2)j_1\ldots j_{p-2k-2}},\pi _{j_1\ldots
j_{p-2k-2}}^{(2)}\right) ,\left( V^{(2)j_1\ldots j_{p-2k-3}},\Pi _{j_1\ldots
j_{p-2k-3}}^{(2)}\right) , 
\end{equation}
with $k\geq 0$. In addition, we set the constraints 
\begin{equation}
\label{s27}\gamma _{i_1\ldots i_{p-2k-2}}^{\prime (1)}\equiv \pi _{i_1\ldots
i_{p-2k-2}}^{(1)}\approx 0,\;k=0,\ldots ,c, 
\end{equation}
\begin{equation}
\label{s28}T_{i_1\ldots i_{p-2k-3}}^{\prime (1)}\equiv \Pi _{i_1\ldots
i_{p-2k-3}}^{(1)}\approx 0,\;k=0,\ldots ,d, 
\end{equation}
\begin{equation}
\label{s29}\gamma _{i_1\ldots i_{p-2k-2}}^{(2)}\equiv -\pi _{i_1\ldots
i_{p-2k-2}}^{(2)}\approx 0,\;k=0,\ldots ,c, 
\end{equation}
\begin{equation}
\label{s30}T_{i_1\ldots i_{p-2k-3}}^{(2)}\equiv -\Pi _{i_1\ldots
i_{p-2k-3}}^{(2)}\approx 0,\;k=0,\ldots ,d, 
\end{equation}
where 
\begin{equation}
\label{s30a}d=\left\{ 
\begin{array}{c}
\frac p2-2,\; 
{\rm for}\;p\;{\rm even}, \\ \frac{p-3}2,\;{\rm for}\;p\;{\rm odd}. 
\end{array}
\right. 
\end{equation}
It is well-known that one can always redefine the surface of first-class
constraints up to a linear combination of constraints whose coefficients
form an invertible matrix. In this respect, we remark that the canonical
momenta $\left( \pi _{i_1\ldots i_{p-2k-2}}\right) _{k=0,\ldots ,c}$ and $%
\left( \Pi _{i_1\ldots i_{p-2k-3}}\right) _{k=0,\ldots ,d}$ can be expressed
with the help of the constraints (\ref{s22}--\ref{s23}) under the form 
\begin{eqnarray}\label{s27a}
& &\pi _{i_1\ldots i_{p-2k-2}}=-\frac 1{M^2+\triangle }\left( \left(
p-2k-1\right) \partial ^i\gamma _{ii_1\ldots i_{p-2k-2}}^{(2)}+MT_{i_1\ldots
i_{p-2k-2}}^{(2)}+\right. \nonumber \\ 
& &\left. \frac 1{p-2k-2}\partial _{\left[ i_1\right. }^{}\gamma
_{\left. i_2\ldots i_{p-2k-2}\right] }^{(2)}\right) , 
\end{eqnarray}
\begin{eqnarray}\label{s28a}
& &\Pi _{i_1\ldots i_{p-2k-3}}=-\frac 1{M^2+\triangle }\left( \left(
p-2k-2\right) \partial ^iT_{ii_1\ldots i_{p-2k-3}}^{(2)}-M\gamma _{i_1\ldots
i_{p-2k-3}}^{(2)}+\right. \nonumber \\
& &\left. \frac 1{p-2k-3}\partial _{\left[ i_1\right. }^{}T_{\left.
i_2\ldots i_{p-2k-3}\right] }^{(2)}\right) . 
\end{eqnarray}
Therefore, we can redefine the constraints (\ref{s27}--\ref{s28}) like 
\begin{equation}
\label{s27b}\gamma _{i_1\ldots i_{p-2k-2}}^{(1)}\equiv \pi _{i_1\ldots
i_{p-2k-2}}-\pi _{i_1\ldots i_{p-2k-2}}^{(1)}\approx 0,\;k=0,\ldots ,c, 
\end{equation}
\begin{equation}
\label{s28b}T_{i_1\ldots i_{p-2k-3}}^{(1)}\equiv \Pi _{i_1\ldots
i_{p-2k-3}}-\Pi _{i_1\ldots i_{p-2k-3}}^{(1)}\approx 0,\;k=0,\ldots ,d. 
\end{equation}
It is clear that the constraints (\ref{s22}--\ref{s23}), (\ref{s20}--\ref
{s21}), (\ref{s29}--\ref{s30}) and (\ref{s27b}--\ref{s28b}) are first-class
and irreducible. The number of physical degrees of freedom of the last
irreducible model coincides with that of the starting reducible theory. The
first-class Hamiltonian corresponding to the theory possessing the above
mentioned irreducible first-class constraints can be chosen of the type 
\begin{eqnarray}\label{s31}
& &H^{\prime }=\bar H+\int d^{d-1}x\left( \sum\limits_{k=0}^cA^{i_1\cdots
i_{p-2k-2}}\pi _{i_1\ldots i_{p-2k-2}}^{(2)}+\sum\limits_{k=0}^dH^{i_1\cdots
i_{p-2k-3}}\Pi _{i_1\ldots i_{p-2k-3}}^{(2)}+\right. \nonumber \\
& &\sum\limits_{k=0}^cB^{(2)i_1\cdots i_{p-2k-2}}\left( \left( p-2k-1\right)
\partial ^i\gamma _{ii_1\ldots i_{p-2k-2}}^{(2)}+MT_{i_1\ldots
i_{p-2k-2}}^{(2)}+\right. \nonumber \\
& &\left. \frac 1{p-2k-2}\partial _{\left[ i_1\right. }^{}\gamma _{\left.
i_2\ldots i_{p-2k-2}\right] }^{(2)}\right)
+\sum\limits_{k=0}^dV^{(2)i_1\cdots i_{p-2k-3}}\left( \left( p-2k-2\right)
\partial ^iT_{ii_1\ldots i_{p-2k-3}}^{(2)}-\right. \nonumber \\
& &\left. \left. M\gamma _{i_1\ldots i_{p-2k-3}}^{(2)}+\frac
1{p-2k-3}\partial _{\left[ i_1\right. }^{}T_{\left. i_2\ldots
i_{p-2k-3}\right] }^{(2)}\right) \right) \equiv \int d^{d-1}xh^{\prime }. 
\end{eqnarray}
In conclusion, starting from action (\ref{s1}) we derived an irreducible
theory based on the first-class constraints (\ref{s22}--\ref{s23}), (\ref
{s20}--\ref{s21}), (\ref{s29}--\ref{s30}), (\ref{s27b}--\ref{s28b}) and on
the first-class Hamiltonian (\ref{s31}). We remark that the above
first-class constraints and first-class Hamiltonian density are local
functions.

\subsection{Irreducible BRST symmetry}

Here we point out the construction of the irreducible BRST symmetry for the
irreducible theory built previously. The minimal antighost spectrum of the
irreducible Koszul-Tate differential is organized as 
\begin{equation}
\label{s32a}\left( {\cal P}_{1i_1\cdots i_{p-2k-1}},\;{\cal P}_{2i_1\cdots
i_{p-2k-1}}\right) ,\;k=0,\ldots ,a, 
\end{equation}
(associated with (\ref{s20}), respectively, (\ref{s22})), 
\begin{equation}
\label{s33a}\left( {\cal P}_{1i_1\cdots i_{p-2k-2}},\;{\cal P}_{2i_1\cdots
i_{p-2k-2}}\right) ,\;k=0,\ldots ,c, 
\end{equation}
(associated with (\ref{s27b}), respectively, (\ref{s29})), 
\begin{equation}
\label{s34a}\left( P_{1i_1\ldots i_{p-2k-2}},\;P_{2i_1\ldots
i_{p-2k-2}}\right) ,\;k=0,\ldots ,c, 
\end{equation}
(corresponding to (\ref{s21}) and (\ref{s23})), plus 
\begin{equation}
\label{s35a}\left( P_{1i_1\ldots i_{p-2k-3}},\;P_{2i_1\ldots
i_{p-2k-3}}\right) ,\;k=0,\ldots ,d, 
\end{equation}
(corresponding to (\ref{s28b}) and (\ref{s30})). All the previous fields are
fermionic, with the ${\cal P}$'s and $P$'s of antighost number one. The
usual definitions of $\delta $ are given by 
\begin{equation}
\label{s35b}\delta z^A=0, 
\end{equation}
\begin{equation}
\label{s35c}\delta {\cal P}_{\Delta i_1\cdots i_{p-2k-1}}=-\gamma
_{i_1\cdots i_{p-2k-1}}^{(\Delta )},\;\Delta =1,2,\;k=0,\ldots ,a, 
\end{equation}
\begin{equation}
\label{s35d}\delta {\cal P}_{\Delta i_1\cdots i_{p-2k-2}}=-\gamma
_{i_1\cdots i_{p-2k-2}}^{(\Delta )},\;\Delta =1,2,\;k=0,\ldots ,c, 
\end{equation}
\begin{equation}
\label{s35e}\delta P_{\Delta i_1\ldots i_{p-2k-2}}=-T_{i_1\cdots
i_{p-2k-2}}^{(\Delta )},\;\Delta =1,2,\;k=0,\ldots ,c, 
\end{equation}
\begin{equation}
\label{s35f}\delta P_{\Delta i_1\ldots i_{p-2k-3}}=-T_{i_1\cdots
i_{p-2k-3}}^{(\Delta )},\;\Delta =1,2,\;k=0,\ldots ,d, 
\end{equation}
where $z^A$ generically denotes any original field/momentum or new variable
in (\ref{s16}), (\ref{s18}) or (\ref{s25}--\ref{s26}). With the help of
these definitions, $\delta $ is found nilpotent and acyclic. The other
differential involved with the BRST symmetry, namely, the longitudinal
derivative along the gauge orbits, requires the minimal ghost spectrum 
\begin{equation}
\label{s32}\left( \eta _1^{i_1\cdots i_{p-2k-1}},\;\eta _2^{i_1\cdots
i_{p-2k-1}}\right) ,\;k=0,\ldots ,a, 
\end{equation}
(associated with (\ref{s20}), respectively, (\ref{s22})), 
\begin{equation}
\label{s33}\left( \eta _1^{i_1\cdots i_{p-2k-2}},\;\eta _2^{i_1\cdots
i_{p-2k-2}}\right) ,\;k=0,\ldots ,c, 
\end{equation}
(associated with (\ref{s27b}), respectively, (\ref{s29})), 
\begin{equation}
\label{s34}\left( C_1^{i_1\ldots i_{p-2k-2}},\;C_2^{i_1\ldots
i_{p-2k-2}}\right) ,\;k=0,\ldots ,c, 
\end{equation}
(corresponding to (\ref{s21}) and (\ref{s23})), and 
\begin{equation}
\label{s35}\left( C_1^{i_1\ldots i_{p-2k-3}},\;C_2^{i_1\ldots
i_{p-2k-3}}\right) ,\;k=0,\ldots ,d, 
\end{equation}
(corresponding to (\ref{s28b}) and (\ref{s30})). The above fields are
fermionic and have the pure ghost number equal to one. The definitions of
the longitudinal derivative along the gauge orbits, $\sigma $, read as%
\begin{eqnarray}\label{s36a}
& &\sigma F=
\sum\limits_{\Delta =1}^2\left( \sum\limits_{k=0}^a\left[ F,\gamma
_{i_1\cdots i_{p-2k-1}}^{(\Delta )}\right] \eta _\Delta ^{i_1\cdots
i_{p-2k-1}}+\sum\limits_{k=0}^c\left[ F,\gamma _{i_1\cdots
i_{p-2k-2}}^{(\Delta )}\right] \eta _\Delta ^{i_1\cdots i_{p-2k-2}}+\right. 
\nonumber \\
& &\left. \sum\limits_{k=0}^a\left[ F,T_{i_1\cdots
i_{p-2k-2}}^{(\Delta )}\right] C_\Delta ^{i_1\cdots
i_{p-2k-2}}+\sum\limits_{k=0}^d\left[ F,T_{i_1\cdots i_{p-2k-3}}^{(\Delta
)}\right] C_\Delta ^{i_1\cdots i_{p-2k-3}}\right) , 
\end{eqnarray}
\begin{equation}
\label{s36b}\sigma {\cal G}^\Gamma =0, 
\end{equation}
where $F$ is any function involving the original or newly added bosonic
canonical pairs, and ${\cal G}^\Gamma $ denote the minimal ghosts (\ref{s32}%
--\ref{s35}). The operator $\sigma $ is in this case strongly nilpotent.
Extending $\sigma $ to the antighosts (\ref{s32a}--\ref{s35a}), generically
denoted by ${\cal P}_\Gamma $, through 
\begin{equation}
\label{s36c}\sigma {\cal P}_\Gamma =0, 
\end{equation}
and $\delta $ to the ghosts (\ref{s32}--\ref{s35}) by means of 
\begin{equation}
\label{s36d}\delta {\cal G}^\Gamma =0, 
\end{equation}
the homological perturbation theory \cite{hom1} ensures that the irreducible
BRST symmetry $s_I=\delta +\sigma $ exists and is nilpotent, $s_I^2=0$. In
conclusion, at this stage we constructed an irreducible BRST symmetry
corresponding to the original reducible one. In the following we find its
relationship with the standard reducible Hamiltonian BRST symmetry of the
starting model.

\subsection{Physical observables}

Here, we establish the link between the reducible and irreducible BRST
symmetries. In this light, we prove that the physical observables
corresponding to the reducible, respectively, irreducible models coincide.
We indicate below the line for $p$ even, the other situation being
investigated in a similar way. Initially, we show that any observable
associated with the irreducible theory is also an observable of the
reducible system. Let $F$ be an observable of the irreducible model. Then,
it fulfils the equations 
\begin{equation}
\label{s23m}\left[ F,\gamma _{i_1\ldots i_{p-2k-1}}^{(1)}\right] \approx
0,\;k=0,\cdots ,a,\;\left[ F,T_{i_1\ldots i_{p-2k-2}}^{(1)}\right] \approx
0,\;k=0,\cdots ,c 
\end{equation}
\begin{equation}
\label{s23n}\left[ F,\gamma _{i_1\ldots i_{p-2k-1}}^{(2)}\right] \approx
0,\;k=0,\cdots ,a, 
\end{equation}
\begin{equation}
\label{s23o}\left[ F,T_{i_1\ldots i_{p-2k-2}}^{(2)}\right] \approx
0,\;k=0,\cdots ,c, 
\end{equation}
\begin{equation}
\label{s28bc}\left[ F,\gamma _{i_1\ldots i_{p-2k-2}}^{(1)}\right] \approx
0,\;\left[ F,\gamma _{i_1\ldots i_{p-2k-2}}^{(2)}\right] \approx
0,\;k=0,\cdots ,c, 
\end{equation}
\begin{equation}
\label{s28bd}\left[ F,T_{i_1\ldots i_{p-2k-3}}^{(1)}\right] \approx
0,\;\left[ F,T_{i_1\ldots i_{p-2k-3}}^{(2)}\right] \approx 0,\;k=0,\cdots
,d. 
\end{equation}
Equations (\ref{s23m}) imply that $F$ does not depend (at least weakly) on
any $A^{0i_1\ldots i_{p-2k-1}}$ or $H^{0i_1\ldots i_{p-2k-2}}$, while (\ref
{s28bc}--\ref{s28bd}) ensure that $F$ does not involve (at least weakly) any
of the fields $B^{(1)j_1\ldots j_{p-2k-2}}$, $B^{(2)j_1\ldots j_{p-2k-2}}$, $%
V^{(1)j_1\ldots j_{p-2k-3}}$ or $V^{(2)j_1\ldots j_{p-2k-3}}$. Next, we
explore the relations (\ref{s23n}--\ref{s23o}). We start from the last
equations in (\ref{s23n}--\ref{s23o}) assuming that $p$ is even 
\begin{equation}
\label{s23p}-2\partial _y^j\left[ F\left( x\right) ,\pi _{ji}\left( y\right)
\right] +M\left[ F\left( x\right) ,\Pi _i\left( y\right) \right] -\partial
_i^y\left[ F\left( x\right) ,\pi \left( y\right) \right] \approx 0, 
\end{equation}
\begin{equation}
\label{s23q}-\partial _y^j\left[ F\left( x\right) ,\Pi _j\left( y\right)
\right] -M\left[ F\left( x\right) ,\pi \left( y\right) \right] \approx 0. 
\end{equation}
Applying $\partial _y^i$ on (\ref{s23p}), multiplying (\ref{s23q}) by $M$
and adding the resulting equations, we arrive at $\left( \partial
_j^y\partial _y^j+M^2\right) \left[ F\left( x\right) ,\pi \left( y\right)
\right] \approx 0$, which yields 
\begin{equation}
\label{s23r}\left[ F\left( x\right) ,\pi \left( y\right) \right] \approx 0. 
\end{equation}
Replacing (\ref{s23r}) back in (\ref{s23p}--\ref{s23q}), these equations
turn into 
\begin{equation}
\label{s23s}-2\partial _y^j\left[ F\left( x\right) ,\pi _{ji}\left( y\right)
\right] +M\left[ F\left( x\right) ,\Pi _i\left( y\right) \right] \approx 0, 
\end{equation}
\begin{equation}
\label{s23t}-\partial _y^j\left[ F\left( x\right) ,\Pi _j\left( y\right)
\right] \approx 0. 
\end{equation}
Taking into account the next equation from (\ref{s23o})%
\begin{eqnarray}\label{s23u}
& &-3\partial _y^l\left[ F\left( x\right) ,\Pi _{lij}\left( y\right) \right]
-M\left[ F\left( x\right) ,\pi _{ij}\left( y\right) \right] -\nonumber \\ 
& &\frac 12\left( \partial _i^y\left[ F\left( x\right) ,\Pi
_j\left( y\right) \right] -\partial _j^y\left[ F\left( x\right) ,\Pi
_i\left( y\right) \right] \right) \approx 0,
\end{eqnarray}
multiplied by $\partial _y^i$ and employing (\ref{s23s}--\ref{s23t}), we get 
$\left( \partial _i^y\partial _y^i+M^2\right) \left[ F\left( x\right) ,\Pi
_j\left( y\right) \right] \approx 0$, hence 
\begin{equation}
\label{s23v}\left[ F\left( x\right) ,\Pi _j\left( y\right) \right] \approx
0. 
\end{equation}
Substituting the result (\ref{s23v}) in (\ref{s23s}) and (\ref{s23u}), it
follows 
\begin{equation}
\label{s23x}-2\partial _y^j\left[ F\left( x\right) ,\pi _{ji}\left( y\right)
\right] \approx 0, 
\end{equation}
\begin{equation}
\label{s23y}-3\partial _y^l\left[ F\left( x\right) ,\Pi _{lij}\left(
y\right) \right] -M\left[ F\left( x\right) ,\pi _{ij}\left( y\right) \right]
\approx 0. 
\end{equation}
Writing down the next equation from (\ref{s23n})%
\begin{eqnarray}\label{s23w}
& &-4\partial _y^l\left[ F\left( x\right) ,\pi _{lijk}\left( y\right) \right] -
M\left[ F\left( x\right) ,\Pi _{ijk}\left( y\right) \right] -
\frac 13\partial _i^y\left[ F\left( x\right) ,
\pi _{jk}\left( y\right) \right] -\nonumber \\ 
& &\frac 13\left( \partial _k^y\left[ F\left( x\right) ,
\pi _{ij}\left( y\right) \right] +\partial _j^y\left[ F\left( x\right) ,
\pi _{ki}\left( y\right) \right] \right) \approx 0,
\end{eqnarray}
on which we apply $\partial _y^i$, and subsequently making use of (\ref{s23x}%
--\ref{s23y}), we finally deduce that $\left( \partial _i^y\partial
_y^i+M^2\right) \left[ F\left( x\right) ,\pi _{jk}\left( y\right) \right]
\approx 0$, which then implies 
\begin{equation}
\label{s23z}\left[ F\left( x\right) ,\pi _{jk}\left( y\right) \right]
\approx 0. 
\end{equation}
Inserting the previous result in (\ref{s23y}--\ref{s23w}) and going on with
the procedure described above, we obtain 
\begin{equation}
\label{s23za}\left[ F\left( x\right) ,\pi _{i_1\cdots i_{p-2k}}\right]
\approx 0,\;k=1,\cdots ,b, 
\end{equation}
\begin{equation}
\label{s23zb}\left[ F\left( x\right) ,\Pi _{i_1\cdots i_{p-2k-1}}\right]
\approx 0,\;k=1,\cdots ,a, 
\end{equation}
which, replaced in the first equations from (\ref{s23n}--\ref{s23o})
(corresponding to $k=0$), lead to 
\begin{equation}
\label{s23zc}\left[ F,G_{i_1\cdots i_{p-1}}^{(2)}\right] \approx 0, 
\end{equation}
\begin{equation}
\label{s23zd}\left[ F,G_{i_1\cdots i_{p-2}}^{(2)}\right] \approx 0. 
\end{equation}
In (\ref{s23za}) we employed the notation 
\begin{equation}
\label{x}b=\left\{ 
\begin{array}{c}
\frac p2,\; 
{\rm for}\;p\;{\rm even}, \\ \frac{p-1}2,\;{\rm for}\;p\;{\rm odd}. 
\end{array}
\right. 
\end{equation}
The equations (\ref{s23za}--\ref{s23zb}) show that $F$ does not depend, at
least weakly, on the fields $A^{i_1\cdots i_{p-2k}}$ and $H^{i_1\cdots
i_{p-2k-1}}$, with $k\geq 1$. As a consequence of our analysis, we managed
to show that an observable $F$ of the irreducible model does not depend (at
least weakly) on the fields $\left( A^{0i_1\ldots i_{p-2k-1}},H^{0i_1\ldots
i_{p-2k-2}}\right) _{k\geq 1}$ (see (\ref{s23m}) with $k\geq 1$), $\left(
A^{i_1\cdots i_{p-2k}},H^{i_1\cdots i_{p-2k-1}}\right) _{k\geq 1}$ (see (\ref
{s23za}--\ref{s23zb})), as well as on $\left( B^{(1)j_1\ldots
j_{p-2k-2}},B^{(2)j_1\ldots j_{p-2k-2}}\right) _{k\geq 0}$, $\left(
V^{(1)j_1\ldots j_{p-2k-3}},V^{(2)j_1\ldots j_{p-2k-3}}\right) _{k\geq 0}$
(see (\ref{s28bc}--\ref{s28bd})) and, in addition, it satisfies the
equations 
\begin{equation}
\label{s23ze}\left[ F,\gamma _{i_1\ldots i_{p-1}}^{(1)}\right] \approx
0,\;\left[ F,T_{i_1\ldots i_{p-2}}^{(1)}\right] \approx 0, 
\end{equation}
(see (\ref{s23m}) with $k=0$) and (\ref{s23zc}--\ref{s23zd}), which are
nothing but the equations verified by an observable of the redundant theory.
Thus, we can conclude that any observable corresponding to the irreducible
system stands for an observable of the original reducible model. The
converse also holds, namely any observable of the redundant system remains
so for the irreducible theory. This is because an observable $\bar F$ of the
original system checks the equations (\ref{s23zc}--\ref{s23zd}), (\ref{s23ze}%
) and does not depend on the newly introduced canonical pairs, such that (%
\ref{s23m}-\ref{s28bd}) are automatically verified. In consequence, the two
theories (reducible and irreducible) possess the same observables, such that
the zeroth order cohomological groups of $s_R$ and $s_I$ coincide 
\begin{equation}
\label{s23abc}H^0\left( s_R\right) =H^0\left( s_I\right) . 
\end{equation}
Thus, the irreducible and reducible theories are equivalent from the BRST
formalism point of view, i.e., from the point of view of the basic equations
underlying the BRST symmetry, $s^2=0$ and $H^0\left( s\right) =\left\{ {\rm %
physical\;observables}\right\} $. This consideration yields the conclusion
that we can replace the BRST quantization of the reducible model by that of
the irreducible theory.

\subsection{Irreducible path integral}

Based on the last conclusion, we approach the Hamiltonian BRST quantization
of the irreducible theory. The minimal antighost and ghost spectra are given
in (\ref{s32a}--\ref{s35a}), respectively, (\ref{s32}--\ref{s35}). In
addition, we further introduce the non-minimal sector 
\begin{equation}
\label{s36}\left( P_{\bar \eta }^{i_1\ldots i_{p-2k-1}},\bar \eta
_{i_1\ldots i_{p-2k-1}}\right) ,\;\left( P_{\bar \eta ^1}^{i_1\ldots
i_{p-2k-1}},\bar \eta _{i_1\ldots i_{p-2k-1}}^1\right) ,\;k=0,\ldots ,a, 
\end{equation}
\begin{equation}
\label{s37}\left( P_b^{i_1\ldots i_{p-2k-1}},b_{i_1\ldots i_{p-2k-1}}\right)
,\;\left( P_{b^1}^{i_1\ldots i_{p-2k-1}},b_{i_1\ldots i_{p-2k-1}}^1\right)
,\;k=0,\ldots ,a, 
\end{equation}
\begin{equation}
\label{s38}\left( P_{\bar C}^{i_1\ldots i_{p-2k-2}},\bar C_{i_1\ldots
i_{p-2k-2}}\right) ,\;\left( P_{\bar C^1}^{i_1\ldots i_{p-2k-2}},\bar
C_{i_1\ldots i_{p-2k-2}}^1\right) ,\;k=0,\ldots ,c, 
\end{equation}
\begin{equation}
\label{s39}\left( P_{\tilde b}^{i_1\ldots i_{p-2k-2}},\tilde b_{i_1\ldots
i_{p-2k-2}}\right) ,\;\left( P_{\tilde b^1}^{i_1\ldots i_{p-2k-2}},\tilde
b_{i_1\ldots i_{p-2k-2}}^1\right) ,\;k=0,\ldots ,c, 
\end{equation}
The fields (\ref{s37}), (\ref{s39}) are all bosonic and possess ghost number
zero, while (\ref{s36}), (\ref{s38}) are fermionic, the $P$'s having ghost
number one, and the $\bar \eta $'s and $\bar C$'s displaying ghost number
minus one. The ghost number is defined as the difference between the pure
ghost number and the antighost number. The non-minimal BRST charge,
respectively, the BRST-invariant extension of $H^{\prime }$ will
consequently be expressed by%
\begin{eqnarray}\label{s40}
& &\Omega =\int d^{d-1}x\left( \sum_{\Delta =1}^2\left( \sum_{k=1}^p\eta
_\Delta ^{i_1\ldots i_{p-k}}\gamma _{i_1\ldots i_{p-k}}^{(\Delta
)}+\sum_{k=1}^{p-1}C_\Delta ^{i_1\ldots i_{p-k-1}}T_{i_1\ldots
i_{p-k-1}}^{(\Delta )}\right) +\right. \nonumber \\
& &\sum_{k=0}^a\left( P_{\bar \eta }^{i_1\ldots i_{p-2k-1}}b_{i_1\ldots
i_{p-2k-1}}+P_{\bar \eta ^1}^{i_1\ldots i_{p-2k-1}}b_{i_1\ldots
i_{p-2k-1}}^1\right) +\nonumber \\ 
& &\left. \sum_{k=0}^c\left( P_{\bar C}^{i_1\ldots i_{p-2k-2}}\tilde
b_{i_1\ldots i_{p-2k-2}}+P_{\bar C^1}^{i_1\ldots i_{p-2k-2}}\tilde
b_{i_1\ldots i_{p-2k-2}}^1\right) \right) , 
\end{eqnarray}
\begin{eqnarray}\label{s41}
& &H_B^{\prime }=
H^{\prime }+\int d^{d-1}x\left( \sum_{k=0}^a\eta _1^{i_1\ldots
i_{p-2k-1}}{\cal P}_{2i_1\ldots i_{p-2k-1}}+\sum_{k=0}^cC_1^{i_1\ldots
i_{p-2k-2}}P_{2i_1\ldots i_{p-2k-2}}-\right. \nonumber \\
& &\sum_{k=0}^c\eta _1^{i_1\ldots i_{p-2k-2}}{\cal P}_{2i_1\ldots
i_{p-2k-2}}-\sum_{k=0}^dC_1^{i_1\ldots i_{p-2k-3}}P_{2i_1\ldots i_{p-2k-3}}+
\nonumber \\
& &\frac 1{p-1}\eta _2^{i_1\ldots i_{p-1}}\partial _{\left[ i_1\right. }%
{\cal P}_{2\left. i_2\ldots i_{p-1}\right] }+\sum_{k=1}^a\eta _2^{i_1\ldots
i_{p-2k-1}}\left( \left( p-2k\right) \partial ^i{\cal P}_{2ii_1\ldots
i_{p-2k-1}}-\right. \nonumber \\
& &\left. MP_{2i_1\ldots i_{p-2k-1}}+\frac 1{p-2k-1}\partial _{\left[
i_1\right. }{\cal P}_{2\left. i_2\ldots i_{p-2k-1}\right] }\right)
+C_2^{i_1\ldots i_{p-2}}\left( M{\cal P}_{2i_1\ldots i_{p-2}}+\right. 
\nonumber \\
& &\left. \frac 1{p-2}\partial _{\left[ i_1\right. }P_{2\left. i_2\ldots
i_{p-2}\right] }\right) +\sum_{k=1}^cC_2^{i_1\ldots i_{p-2k-2}}\left( \left(
p-2k-1\right) \partial ^iP_{2ii_1\ldots i_{p-2k-2}}+\right. \nonumber \\ 
& &\left. M{\cal P}_{2i_1\ldots i_{p-2k-2}}+\frac 1{p-2k-2}\partial _{\left[
i_1\right. }P_{2\left. i_2\ldots i_{p-2k-2}\right] }\right) -\nonumber \\
& &\sum_{k=0}^c\eta _2^{i_1\ldots i_{p-2k-2}}\left( \left( p-2k-1\right)
\partial ^i{\cal P}_{2ii_1\ldots i_{p-2k-2}}+MP_{2i_1\ldots
i_{p-2k-2}}+\right. \nonumber \\
& &\left. +\frac 1{p-2k-2}\partial _{\left[ i_1\right. }{\cal P}_{2\left.
i_2\ldots i_{p-2k-2}\right] }\right) -\sum_{k=0}^dC_2^{i_1\ldots
i_{p-2k-3}}\left( \left( p-2k-2\right) \partial ^iP_{2ii_1\ldots
i_{p-2k-3}}-\right. \nonumber \\
& &\left. \left. M{\cal P}_{2i_1\ldots i_{p-2k-3}}+\frac
1{p-2k-3}\partial _{\left[ i_1\right. }P_{2\left. i_2\ldots
i_{p-2k-3}\right] }\right) \right) . 
\end{eqnarray}
We take the gauge-fixing fermion%
\begin{eqnarray}\label{s42}
& &K=\int d^{d-1}x\left( {\cal P}_{1i_1\ldots i_{p-1}}\left( \partial
_iA^{ii_1\ldots i_{p-1}}+MH^{i_1\ldots i_{p-1}}+\partial ^{\left[ i_1\right.
}B^{(1)\left. i_2\ldots i_{p-1}\right] }\right) +\right. \nonumber \\
& &\sum_{k=1}^a{\cal P}_{1i_1\ldots i_{p-2k-1}}\left( \partial
_iB^{(1)ii_1\ldots i_{p-2k-1}}+MV^{(1)i_1\ldots i_{p-2k-1}}+\partial
^{\left[ i_1\right. }B^{(1)\left. i_2\ldots i_{p-2k-1}\right] }\right) + 
\nonumber \\
& &P_{1i_1\ldots i_{p-2}}\left( \partial _iH^{ii_1\ldots
i_{p-2}}-MB^{(1)i_1\ldots i_{p-2}}+\partial ^{\left[ i_1\right.
}V^{(1)\left. i_2\ldots i_{p-2}\right] }\right) +\nonumber \\ 
& &\sum_{k=1}^cP_{1i_1\ldots i_{p-2k-2}}\left( \partial _iV^{(1)ii_1\ldots
i_{p-2k-2}}-MB^{(1)i_1\ldots i_{p-2k-2}}+\partial ^{\left[ i_1\right.
}V^{(1)\left. i_2\ldots i_{p-2k-2}\right] }\right) +\nonumber \\ 
& &\left( -\right) ^{p+1}\sum_{k=0}^c{\cal P}_{1i_1\ldots i_{p-2k-2}}\left(
\partial _iA^{ii_1\ldots i_{p-2k-2}0}+MH^{i_1\ldots i_{p-2k-2}0}+\right.
\nonumber \\
& &\left. \partial ^{\left[ i_1\right. }A^{\left. i_2\ldots i_{p-2k-2}\right]
0}\right) +\left( -\right) ^p\sum_{k=0}^dP_{1i_1\ldots i_{p-2k-3}}\left(
\partial _iH^{ii_1\ldots i_{p-2k-3}0}-\right. \nonumber \\
& &\left. MA^{i_1\ldots i_{p-2k-3}0}+\partial ^{\left[ i_1\right. }H^{\left.
i_2\ldots i_{p-2k-3}\right] 0}\right) +\nonumber \\
& &\sum_{k=0}^aP_{b^1}^{i_1\ldots i_{p-2k-1}}\left( {\cal P}_{1i_1\ldots
i_{p-2k-1}}-\bar \eta _{i_1\ldots i_{p-2k-1}}+\stackrel{.}{\bar \eta }%
_{i_1\ldots i_{p-2k-1}}^1\right) +\nonumber \\
& &\sum_{k=0}^aP_b^{i_1\ldots i_{p-2k-1}}\left( \bar \eta _{i_1\ldots
i_{p-2k-1}}^1+\stackrel{.}{\bar \eta }_{i_1\ldots i_{p-2k-1}}\right) +
\nonumber \\
& &\sum_{k=0}^cP_{\tilde b}^{i_1\ldots i_{p-2k-2}}\left( \bar C_{i_1\ldots
i_{p-2k-2}}^1+\stackrel{.}{\bar C}_{i_1\ldots i_{p-2k-2}}\right) +\nonumber \\
& &\left. \sum_{k=0}^cP_{\tilde b^1}^{i_1\ldots i_{p-2k-2}}\left(
P_{1i_1\ldots i_{p-2k-2}}-\bar C_{i_1\ldots i_{p-2k-2}}+\stackrel{.}{\bar C}%
_{i_1\ldots i_{p-2k-2}}^1\right) \right) , 
\end{eqnarray}
and obtain after some computation the path integral%
\begin{eqnarray}\label{s43}
& &Z_K=\int {\cal D}A^{\mu _1\ldots \mu _p}DH^{\mu _1\ldots \mu _{p-1}}\left(
\prod_{k=0}^c{\cal D}B^{(1)\mu _1\ldots \mu _{p-2k-2}}\right) \left(
\prod_{k=0}^d{\cal D}V^{(1)\mu _1\ldots \mu _{p-2k-3}}\right) \times 
\nonumber \\
& &\left( \prod_{k=0}^c\left( {\cal D}\tilde b_{\mu _1\ldots \mu _{p-2k-2}}%
{\cal D}C_2^{\mu _1\ldots \mu _{p-2k-2}}{\cal D}\bar C_{\mu _1\ldots \mu
_{p-2k-2}}\right) \right) \times \nonumber \\
& &\left( \prod_{k=0}^a\left( {\cal D}b_{\mu _1\ldots \mu _{p-2k-1}}%
{\cal D}\eta _2^{\mu _1\ldots \mu _{p-2k-1}}{\cal D}\bar \eta _{\mu _1\ldots
\mu _{p-2k-1}}\right) \right) \exp iS_K, 
\end{eqnarray}
where 
\begin{eqnarray}\label{s44}
& &S_K=S_0^L+\int d^dx\left( -\sum_{k=0}^a\bar \eta _{\mu _1\ldots \mu
_{p-2k-1}}\left( \Box +M^2\right) \eta _2^{\mu _1\ldots \mu
_{p-2k-1}}-\right. \nonumber \\ 
& &\sum_{k=0}^c\bar C_{\mu _1\ldots \mu _{p-2k-2}}\left( \Box +M^2\right)
C_2^{\mu _1\ldots \mu _{p-2k-2}}+\nonumber \\
& &b_{\mu _1\ldots \mu _{p-1}}\left( \partial _\mu A^{\mu \mu _1\ldots \mu
_{p-1}}+MH^{\mu _1\ldots \mu _{p-1}}+\partial ^{\left[ \mu _1\right.
}B^{(1)\left. \mu _2\ldots \mu _{p-1}\right] }\right) +\nonumber \\ 
& &\sum_{k=1}^ab_{\mu _1\ldots \mu _{p-2k-1}}%
\left( \partial _\mu B^{(1)\mu \mu
_1\ldots \mu _{p-2k-1}}+MV^{(1)\mu _1\ldots \mu _{p-2k-1}}\right. 
\nonumber \\
& &\left. +\partial ^{\left[ \mu _1\right. }B^{(1)\left. \mu _2\ldots \mu
_{p-2k-1}\right] }\right) +\tilde b_{\mu _1\ldots \mu _{p-2}}\left( \partial
_\mu H^{\mu \mu _1\ldots \mu _{p-2}}-\right. \nonumber \\
& &\left. MB^{(1)\mu _1\ldots \mu _{p-2}}+\partial ^{\left[ \mu _1\right.
}V^{(1)\left. \mu _2\ldots \mu _{p-2}\right] }\right) +\nonumber \\ 
& &\sum_{k=1}^c\tilde b_{\mu _1\ldots \mu _{p-2k-2}}\left( \partial _\mu
V^{(1)\mu \mu _1\ldots \mu _{p-2k-2}}-MB^{(1)\mu _1\ldots \mu
_{p-2k-2}}+\right. \nonumber \\
& &\left. \left. \partial ^{\left[ \mu _1\right. }V^{(1)\left. \mu
_2\ldots \mu _{p-2k-2}\right] }\right) \right) ,
\end{eqnarray}
and $S_0^L$ is given by (\ref{s1}). We remark that in deriving (\ref{s44})
we realized the identifications 
\begin{equation}
\label{s46}B^{(1)\mu _1\ldots \mu _{p-2k-2}}\equiv \left( A^{0i_1\ldots
i_{p-2k-3}},B^{(1)i_1\ldots i_{p-2k-2}}\right) ,\;k=0,\ldots ,c, 
\end{equation}
\begin{equation}
\label{s47}V^{(1)\mu _1\ldots \mu _{p-2k-3}}\equiv \left( H^{0i_1\ldots
i_{p-2k-4}},V^{(1)i_1\ldots i_{p-2k-3}}\right) ,\;k=0,\ldots ,d, 
\end{equation}
\begin{equation}
\label{s48}b_{\mu _1\ldots \mu _{p-2k-1}}\equiv \left( \left( p-2k-1\right)
\pi _{i_1\ldots i_{p-2k-2}}^{(1)},b_{i_1\ldots i_{p-2k-1}}\right)
,\;k=0,\ldots ,a, 
\end{equation}
\begin{equation}
\label{s49}\tilde b_{\mu _1\ldots \mu _{p-2k-2}}\equiv \left( \left(
p-2k-2\right) \Pi _{i_1\ldots i_{p-2k-3}}^{(1)},\tilde b_{i_1\ldots
i_{p-2k-2}}\right) ,\;k=0,\ldots ,c, 
\end{equation}
\begin{equation}
\label{s50}\eta _2^{\mu _1\ldots \mu _{p-2k-1}}\equiv \left( \eta
_2^{i_1\ldots i_{p-2k-2}},\eta _2^{i_1\ldots i_{p-2k-1}}\right)
,\;k=0,\ldots ,a, 
\end{equation}
\begin{equation}
\label{s51}C_2^{\mu _1\ldots \mu _{p-2k-2}}\equiv \left( C_2^{i_1\ldots
i_{p-2k-3}},C_2^{i_1\ldots i_{p-2k-2}}\right) ,\;k=0,\ldots ,c, 
\end{equation}
\begin{equation}
\label{s52}\bar \eta _{\mu _1\ldots \mu _{p-2k-1}}\equiv \left( -\left(
p-2k-1\right) {\cal P}_{1i_1\ldots i_{p-2k-2}},\bar \eta _{i_1\ldots
i_{p-2k-1}}\right) ,\;k=0,\ldots ,a, 
\end{equation}
\begin{equation}
\label{s53}\bar C_{\mu _1\ldots \mu _{p-2k-2}}\equiv \left( -\left(
p-2k-2\right) P_{1i_1\ldots i_{p-2k-3}},\bar C_{i_1\ldots i_{p-2k-2}}\right)
,\;k=0,\ldots ,c. 
\end{equation}
In conclusion, we succeeded in deriving a gauge-fixed action with no
residual gauge invariances without introducing the ghosts for ghosts, which
moreover, is Lorentz covariant.

\section{Interacting theories with Stueckelberg coupling}

The treatment from Section 2 starts from the Lagrangian action (\ref{s1}),
which is a quadratic action. In this section we study the possibility to
perform an irreducible analysis in connection with interacting theories
displaying Stueckelberg coupling. More precisely, we investigate what
happens to our irreducible approach if we add to (\ref{s1}) some interaction
terms which are invariant under the gauge transformations (\ref{s1a}--\ref
{s1b}). An idea would be to realize the canonical analysis of the
interacting theory and to further apply the irreducible treatment developed
in the above. However, the interaction terms may contain higher order
derivatives of the fields, such that the canonical analysis becomes
intricate. In this context, it appears the question whether there exists a
more direct irreducible method for interacting theories. As it will be seen
in the sequel, the answer is affirmative. In this light, we prove that our
irreducible Hamiltonian formalism induces an irreducible Lagrangian method
that simplifies the approach to interacting theories. In view of this we
investigate the gauge invariances of the Lagrangian action associated with
the irreducible Hamiltonian formulation of abelian $p$- and $\left(
p-1\right) $-forms with Stueckelberg coupling. From the point of view of the
Hamiltonian BRST quantization, the split into primary and secondary
constraints is not important. In this situation, what matters is the
Hamiltonian gauge algebra. On the contrary, in order to obtain the gauge
transformations of the Lagrangian action, it is necessary to distinguish
between the primary and secondary first-class constraints. This is why we
work with a model of Hamiltonian theory in the case of Stueckelberg coupling
for which we assume that (\ref{s20}--\ref{s21}) and (\ref{s27b}--\ref{s28b})
are primary, while (\ref{s22}--\ref{s23}) and (\ref{s29}--\ref{s30}) are
secondary constraints. The corresponding extended action takes the form%
\begin{eqnarray}\label{s65}
& &S_0^{\prime E}=\int d^dx\left( \sum\limits_{k=0}^b\dot A^{\mu _1\cdots \mu
_{p-2k}}\pi _{\mu _1\cdots \mu _{p-2k}}+\sum\limits_{k=0}^a\dot H^{\mu
_1\cdots \mu _{p-2k-1}}\Pi _{\mu _1\cdots \mu _{p-2k-1}}+\right. \nonumber \\
& &\sum\limits_{\Delta =1}^2\sum\limits_{k=0}^c\dot B^{(\Delta )i_1\cdots
i_{p-2k-2}}\pi _{i_1\cdots i_{p-2k-2}}^{(\Delta )}+\sum\limits_{\Delta
=1}^2\sum\limits_{k=0}^d\dot V^{(\Delta )i_1\cdots i_{p-2k-3}}\Pi
_{i_1\cdots i_{p-2k-3}}^{(\Delta )}-h^{\prime }-\nonumber \\ 
& &\sum\limits_{\Delta =1}^2\sum\limits_{k=0}^au^{(\Delta )i_1\cdots
i_{p-2k-1}}\gamma _{i_1\cdots i_{p-2k-1}}^{(\Delta )}-\sum\limits_{\Delta
=1}^2\sum\limits_{k=0}^c\bar u^{(\Delta )i_1\cdots i_{p-2k-2}}T_{i_1\cdots
i_{p-2k-2}}^{(\Delta )}-\nonumber \\ 
& &\left. \sum\limits_{\Delta =1}^2\sum\limits_{k=0}^cu^{(\Delta
)i_1\cdots i_{p-2k-2}}\gamma _{i_1\cdots i_{p-2k-2}}^{(\Delta
)}-\sum\limits_{\Delta =1}^2\sum\limits_{k=0}^d\bar u^{(\Delta )i_1\cdots
i_{p-2k-3}}T_{i_1\cdots i_{p-2k-3}}^{(\Delta )}\right) , 
\end{eqnarray}
where the $u^{(\Delta )}$'s and $\bar u^{(\Delta )}$'s denote the Lagrange
multipliers of the corresponding constraints. From the extended action (\ref
{s65}) we can obtain the so-called total action by setting all the
multipliers of the type $u^{(2)}$ and $\bar u^{(2)}$ equal to zero. On the
other hand, from the gauge transformations of (\ref{s65}) we can determine
those of the total action by also taking the gauge variations of the $%
u^{(2)} $'s and $\bar u^{(2)}$'s equal to zero. Finally, from the gauge
transformations of the total action we consequently arrive at the Lagrangian
gauge invariances by eliminating the momenta and remaining multipliers on
their equations of motion. In this way, starting from (\ref{s65}) we derive
the corresponding Lagrangian gauge transformations 
\begin{equation}
\label{s54}\delta _\epsilon A^{\mu _1\ldots \mu _p}=\partial ^{\left[ \mu
_1\right. }\epsilon ^{\left. \mu _2\ldots \mu _p\right] }, 
\end{equation}
\begin{equation}
\label{s55}\delta _\epsilon B^{(1)\mu _1\ldots \mu _{p-2k-2}}=\partial
^{\left[ \mu _1\right. }\epsilon ^{\left. \mu _2\ldots \mu _{p-2k-2}\right]
}-M\bar \epsilon ^{\mu _1\ldots \mu _{p-2k-2}}+\partial _\mu \epsilon ^{\mu
\mu _1\ldots \mu _{p-2k-2}}, 
\end{equation}
for $k=0,\ldots ,c$, 
\begin{equation}
\label{s56}\delta _\epsilon H^{\mu _1\ldots \mu _{p-1}}=\partial ^{\left[
\mu _1\right. }\bar \epsilon ^{\left. \mu _2\ldots \mu _{p-1}\right]
}+M\epsilon ^{\mu _1\ldots \mu _{p-1}}, 
\end{equation}
\begin{equation}
\label{s57}\delta _\epsilon V^{(1)\mu _1\ldots \mu _{p-2k-3}}=\partial
^{\left[ \mu _1\right. }\bar \epsilon ^{\left. \mu _2\ldots \mu
_{p-2k-3}\right] }+M\epsilon ^{\mu _1\ldots \mu _{p-2k-3}}+\partial _\mu
\bar \epsilon ^{\mu \mu _1\ldots \mu _{p-2k-3}}, 
\end{equation}
for $k=0,\ldots ,d$. The gauge parameters appearing in (\ref{s54}--\ref{s57}%
) are given by 
\begin{equation}
\label{s58}\epsilon ^{\mu _1\ldots \mu _{p-2k-1}}\equiv \left( \epsilon
^{i_1\ldots i_{p-2k-2}},\epsilon ^{i_1\ldots i_{p-2k-1}}\right)
,\;k=0,\cdots ,a, 
\end{equation}
\begin{equation}
\label{s59}\bar \epsilon ^{\mu _1\ldots \mu _{p-2k-2}}\equiv \left( \bar
\epsilon ^{i_1\ldots i_{p-2k-3}},\bar \epsilon ^{i_1\ldots
i_{p-2k-2}}\right) ,\;k=0,\cdots ,c, 
\end{equation}
with $\left( \epsilon ^{i_1\ldots i_{p-2k-2}},\bar \epsilon ^{i_1\ldots
i_{p-2k-3}}\right) $ corresponding to the constraints (\ref{s29}),
respectively, (\ref{s30}), and $\left( \epsilon ^{i_1\ldots i_{p-2k-1}},\bar
\epsilon ^{i_1\ldots i_{p-2k-2}}\right) $ associated with (\ref{s22}),
respectively, (\ref{s23}). The fields $B^{(1)\mu _1\ldots \mu _{p-2k-2}}$
and $V^{(1)\mu _1\ldots \mu _{p-2k-3}}$ are identified with 
\begin{equation}
\label{s66}B^{(1)\mu _1\ldots \mu _{p-2k-2}}\equiv \left( A^{0i_1\ldots
i_{p-2k-3}},B^{(1)i_1\ldots i_{p-2k-2}}\right) ,\;k=0,\cdots ,c, 
\end{equation}
\begin{equation}
\label{s67}V^{(1)\mu _1\ldots \mu _{p-2k-3}}\equiv \left( H^{0i_1\ldots
i_{p-2k-4}},V^{(1)i_1\ldots i_{p-2k-3}}\right) ,\;k=0,\cdots ,d. 
\end{equation}
If we eliminate all the momenta, the Lagrange multipliers, all the fields
carrying the superscript $(2)$ and the fields $\left( A^{i_1\ldots
i_{p-2k-2}}\right) _{k=0,\cdots ,c}$, $\left( H^{i_1\ldots
i_{p-2k-3}}\right) _{k=0,\cdots ,d}$ on their equations of motion, we derive
that the Lagrangian action resulting from (\ref{s65}) is identical with the
original one, i.e., 
\begin{eqnarray}\label{s68}
& &S_0^{\prime L}\left[ A^{\mu _1\ldots \mu _p},H^{\mu _1\ldots \mu
_{p-1}},B^{(1)\mu _1\ldots \mu _{p-2k-2}},V^{(1)\mu _1\ldots \mu
_{p-2k-3}}\right] =\nonumber \\ 
& &S_0^L\left[ A^{\mu _1\ldots \mu _p},H^{\mu _1\ldots \mu
_{p-1}}\right] . 
\end{eqnarray}
The theory based on action (\ref{s68}) and subject to the irreducible gauge
transformations (\ref{s54}--\ref{s57}) represents the Lagrangian
manifestation of the irreducible Hamiltonian model constructed in Section 2.
The crucial feature of this irreducible Lagrangian theory is that it leads
to (\ref{s43}--\ref{s44}) via the antifield BRST formalism by using an
appropriate gauge-fixing fermion. The minimal solution to the master
equation associated with the above irreducible Lagrangian theory is given by 
\begin{eqnarray}\label{s68a}
& &S=S_0^L\left[ A^{\mu _1\ldots \mu _p},H^{\mu _1\ldots \mu _{p-1}}\right]
+\int d^dx\left( A_{\mu _1\ldots \mu _p}^{*}\partial ^{\left[ \mu _1\right.
}\eta ^{\left. \mu _2\ldots \mu _p\right] }+\right. \nonumber \\ 
& &\sum\limits_{k=0}^cB_{\mu _1\ldots \mu _{p-2k-2}}^{(1)*}\left( \partial
^{\left[ \mu _1\right. }\eta ^{\left. \mu _2\ldots \mu _{p-2k-2}\right]
}-MC^{\mu _1\ldots \mu _{p-2k-2}}+\partial _\mu \eta ^{\mu \mu _1\ldots \mu
_{p-2k-2}}\right) +\nonumber \\ 
& &H_{\mu _1\ldots \mu _{p-1}}^{*}\left( \partial ^{\left[ \mu _1\right.
}C^{\left. \mu _2\ldots \mu _{p-1}\right] }+M\eta ^{\mu _1\ldots \mu
_{p-1}}\right) +\nonumber \\ 
& &\sum\limits_{k=0}^dV_{\mu _1\ldots \mu _{p-2k-3}}^{(1)*}\left( \partial
^{\left[ \mu _1\right. }C^{\left. \mu _2\ldots \mu _{p-2k-3}\right] }+M\eta
^{\mu _1\ldots \mu _{p-2k-3}}+\right. \nonumber \\ 
& &\left. \left. \partial _\mu C^{\mu \mu _1\ldots \mu
_{p-2k-3}}\right) \right) , 
\end{eqnarray}
where $\left( \eta ^{\mu _1\ldots \mu _{p-2k-1}}\right) _{k=0,\cdots ,a}$
and $\left( C^{\mu _1\ldots \mu _{p-2k-2}}\right) _{k=0,\cdots ,c}$
represent the Lagrangian ghost number one ghosts, and the star variables
stand for the antifields of the corresponding fields. Taking the non-minimal
solution as 
\begin{eqnarray}\label{s69}
& &S^{\prime }=S-\int d^dx\left( \sum\limits_{k=0}^a\bar \eta _{\mu _1\ldots
\mu _{p-2k-1}}^{*}b^{\mu _1\ldots \mu _{p-2k-1}}+\right. \nonumber \\ 
& &\left. \sum\limits_{k=0}^c\bar C_{\mu _1\ldots \mu
_{p-2k-2}}^{*}\tilde b^{\mu _1\ldots \mu _{p-2k-2}}\right) , 
\end{eqnarray}
and the gauge-fixing fermion of the form 
\begin{eqnarray}\label{s70}
& &\psi =-\int d^dx\left( \sum\limits_{k=0}^a\bar \eta _{\mu _1\ldots \mu
_{p-2k-1}}M^{\mu _1\ldots \mu _{p-2k-1}}+\right. \nonumber \\
& &\left. \sum\limits_{k=0}^c\bar C_{\mu _1\ldots \mu
_{p-2k-2}}N^{\mu _1\ldots \mu _{p-2k-2}}\right) , 
\end{eqnarray}
we get (\ref{s43}--\ref{s44}) modulo the identifications 
\begin{equation}
\label{s70a}\eta ^{\mu _1\ldots \mu _{p-2k-1}}\equiv \eta _2^{\mu _1\ldots
\mu _{p-2k-1}},\;C^{\mu _1\ldots \mu _{p-2k-2}}\equiv C_2^{\mu _1\ldots \mu
_{p-2k-2}}. 
\end{equation}
In (\ref{s70}) the functions $M$ and $N$ read as 
\begin{equation}
\label{s71}M^{\mu _1\ldots \mu _{p-1}}=\partial _\mu A^{\mu \mu _1\ldots \mu
_{p-1}}+MH^{\mu _1\ldots \mu _{p-1}}+\partial ^{\left[ \mu _1\right.
}B^{(1)\left. \mu _2\ldots \mu _{p-1}\right] }, 
\end{equation}
\begin{eqnarray}\label{s72}
& &M^{\mu _1\ldots \mu _{p-2k-1}}=\partial _\mu B^{(1)\mu \mu _1\ldots \mu
_{p-2k-1}}+\nonumber \\
& &MV^{(1)\mu _1\ldots \mu _{p-2k-1}}+\partial ^{\left[ \mu
_1\right. }B^{(1)\left. \mu _2\ldots \mu _{p-2k-1}\right] },\;k\geq 1,
\end{eqnarray}
\begin{equation}
\label{s73}N^{\mu _1\ldots \mu _{p-2}}=\partial _\mu H^{\mu \mu _1\ldots \mu
_{p-2}}-MB^{(1)\mu _1\ldots \mu _{p-2}}+\partial ^{\left[ \mu _1\right.
}V^{(1)\left. \mu _2\ldots \mu _{p-2}\right] }, 
\end{equation}
\begin{eqnarray}\label{s74}
& &N^{\mu _1\ldots \mu _{p-2k-2}}=\partial _\mu V^{(1)\mu \mu _1\ldots \mu
_{p-2k-2}}-\nonumber \\ 
& &MB^{(1)\mu _1\ldots \mu _{p-2k-2}}+\partial ^{\left[ \mu
_1\right. }V^{(1)\left. \mu _2\ldots \mu _{p-2k-2}\right] },\;k\geq 1.
\end{eqnarray}
The fields $\bar \eta ^{\mu _1\ldots \mu _{p-2k-1}}$, $\bar C^{\mu _1\ldots
\mu _{p-2k-2}}$, $b^{\mu _1\ldots \mu _{p-2k-1}}$ and $\tilde b^{\mu
_1\ldots \mu _{p-2k-2}}$ together with the attached antifields, which carry
a star superscript, form the Lagrangian non-minimal sector. Hence, until now
we proved that the irreducible Lagrangian version for the quadratic theory
leads to the same path integral like the Hamiltonian one. Based on this
result, we can simply solve the interacting case. If we add to action (\ref
{s1}) any interaction terms that are gauge invariant under (\ref{s1a}--\ref
{s1b}), the starting point of the irreducible approach is expressed by the
interacting action, which is invariant under the gauge transformations (\ref
{s54}--\ref{s57}). In this situation the non-minimal solution of the master
equation can be obtained from (\ref{s69}) by adding the starting interaction
terms. Using the same gauge-fixing fermion, namely, (\ref{s70}), we reach a
gauge-fixed action that coincides with (\ref{s44}) apart from the starting
Lagrangian action, which must include the gauge-invariant interaction
pieces. This solves the interacting case discussed here.

\section{Conclusion}

To conclude with, in this paper we presented an irreducible BRST approach to
interacting $p$-form gauge theories with Stueckelberg coupling. Our
procedure includes two basic steps. The first one is relying on the
irreducible Hamiltonian analysis of the quadratic action describing
Stueckelberg coupled $p$- and $\left( p-1\right) $-forms. The irreducible
treatment is mainly based on some irreducible first-class constraints
associated with the original reducible ones. The derivation of the
irreducible constraint set is realized by requiring that all the antighost
number one co-cycles of the Koszul-Tate differential identically vanish
under an appropriate redefinition of the antighost number one antighosts,
and, in the meantime, that the number of physical degrees of freedom should
remain unchanged by passing to the irreducible context. The approach to
interacting theories with Stueckelberg coupling is strongly related to the
irreducible method developed for the quadratic action. Thus, beginning with
the Hamiltonian gauge transformations generated by the irreducible
first-class constraints we derive their Lagrangian version, which, in turn,
is a good starting point for an irreducible BRST approach to interacting
theories with Stueckelberg coupling.

\end{document}